\documentclass[10pt,leqno]{amsart}
\usepackage{graphicx}
\usepackage{indentfirst,csquotes}

\usepackage{amssymb,amsthm,amsmath}
\usepackage{xcolor,paralist,hyperref,titlesec,fancyhdr,etoolbox}

\titleformat{\section}[display]{\normalfont\huge\bfseries\centering}{\centering\chaptertitlename\thechapter}{10pt}{\Large}
\titlespacing*{\section}{0pt}{0ex}{0ex}

\hypersetup{ colorlinks=true, linkcolor=black, filecolor=black, urlcolor=black }

\usepackage{lipsum}

\begin{document}
\title{A DYNAMIC COMBUSTION MODEL FOR SUPERSONIC TURBULENT COMBUSTION} 
\author{\textbf{Xu Zhu, Jian An*, Nana Wang, Jian Zhang, Zhuyin Ren}}
\date{\today}
\address{Institute for Aero Engine, Tsinghua University, Beijing, China}
\email{anjian\_aero@foxmail.com}
\maketitle

\begin{abstract}
Supersonic combustion plays a vital role in various applications, including scramjets, dual-mode ramjets, and pulse detonation engines.
However, the flame characteristics can vary significantly, depending on the application. To model supersonic combustion,
large eddy simulation coupled with a partially stirred reactor (PaSR) is commonly used.
This method assumes that reactions occur at turbulent fine flame structure in a computational cell,
making it less effective for flames governed by not only turbulent mixing but also homogenous autoignition,
which is common under supersonic flow. To address this limitation, this study proposes a novel dynamic combustion model
to enable more versatile modeling of supersonic flames. The model utilizes a two-delta probability density function,
which represents the sub-grid composition variation and models the filtered sub-grid reaction rate.
The corresponding weights are dynamically modeled based on the level of cell composition inhomogeneity.
The new dynamic model is tested in a supersonic combustion case study with a strut-cavity flame holder.
Results demonstrate that the new dynamic model can properly recover the limits of supersonic flames that are
primarily governed by homogenous autoignition and turbulent mixing.  
\end{abstract} 

\bigskip

\section*{INTRODUCTION}
To achieve navigation within and beyond the atmosphere, hypersonic flight technology has garnered significant interest. 
At high altitudes and velocities, the scramjet engine can compress air through the ramming action alone, 
achieving a high pressure ratio without requiring rotating components. 
To investigate the underlying mechanisms, researchers have designed sophisticated experimental devices focusing on cavity-based combustors 
\cite{mathur_supersonic_1999,micka_combustion_2009} and strut-based combustors \cite{genin_simulation_2010}. 
However, these studies often yield only basic flame structure and wall static pressure data, 
making it challenging to accurately portray the spatio-temporal evolution of flow fields and thermochemical information within the combustion chamber. 
Consequently, numerical simulations of turbulent combustion have emerged as another crucial method for examining supersonic combustion phenomena.

Large eddy simulation (LES) offers a promising approach for elucidating the complexities of turbulence. 
However, due to the short residence time in hypersonic combustion, 
the mixing time scale closely aligns with the chemical reaction time scale, 
resulting in a strong coupling effect between the two. 
This interdependence makes it challenging to accurately model the source term of the chemical reaction.
Numerous methods have been proposed and applied for modeling turbulence-chemistry interaction in supersonic reacting flows, 
including well-stirred reactor (WSR) models \cite{genin_simulation_2010}, partially stirred reactor (PaSR) models \cite{berglund_finite_2010}, 
quasi-laminar (QL) reaction rate model \cite{fureby2012supersonic}, the eddy dissipation concept (EDC) model \cite{dally1998flow},
flamelet progress variable models \cite{hou_partially_2014}, and transported probability density function (TPDF) models \cite{koo_quadrature-based_2011}. 

The PaSR combustion model, capable of accounting for finite-rate chemistry and the intricate turbulence-chemistry interaction, 
has demonstrated its effectiveness for supersonic combustion. 
It has been successfully applied in LES studies of ONERA \cite{berglund_finite_2010}, DLR \cite{berglund_les_2007}, and HyShot II scramjet combustors \cite{fureby_cfd_2011}. 
As mentioned in \cite{fureby2012supersonic}, the reacting structures in highly turbulent supersonic flames exhibit a correlation with the dissipative structures of the flow, 
which are typically smaller than the filter width $\Delta$. Generally, these flames are highly wrinkled and fragmented, 
with the assumption that they consist of reacting fine structures, while the surroundings are dominated by large-scale structures. As a result, 
each LES cell can be regarded as a partially stirred reactor containing homogeneous fine structures, exchanging mass and energy with the surrounding environment. 
Turbulence combustion interaction in the well-stirred region adheres to the fundamental assumptions of the WSR model, 
while chemical reactions in the reverse structure region occur considerably slower due to molecular diffusion constraints. 
As a result, the filtered chemical reaction source term can be ignored during the solution process. 

However, given that the PaSR model's basic assumption relies on turbulence mixing intensity and quasi-static process approximation, 
it exhibits limitations in addressing ignition or self-ignition phenomena \cite{sabelnikov_extended_2013} with weak coupling relationships to turbulence action. 
Hypersonic combustion processes frequently involve ignition or self-ignition events induced by high temperatures and significant shear. 
Thus, it is essential to extend the PaSR model's capabilities to encompass supersonic combustion scenarios.
This study tackles this challenge by developing a dynamic partially premixed combustion model, 
building upon the original PaSR method to close the chemical reaction source terms in complex and highly transient turbulent combustion for large eddy simulation. 
A numerical simulation is performed on a typical rocket based combined cycle (RBCC) model combustion chamber example, and the results of the dynamic method are compared with those obtained using the original PaSR method.

\section*{METHODOLOGY}
\subsection*{Large Eddy Simulation Theory}

For LES, the flow field value $\phi$ undergoes filtering through a low-pass filtering function $G(x)$, as shown in the following equation:
\begin{equation}
	\bar{\phi}(x, t)=\int_V \phi\left(x^{\prime}, t\right) G\left(x-x^{\prime}\right) d x^{\prime}
\end{equation}

Consequently, the filtered control equations are obtained
\begin{equation}
	\begin{gathered}
		\frac{\partial \bar{\rho}}{\partial t}+\frac{\partial \bar{\rho} \tilde{u}_j}{\partial x_j}=0 \\
		\frac{\partial}{\partial t}\left(\bar{\rho} \tilde{u}_i\right)+\frac{\partial}{\partial x_j}\left(\bar{\rho} \tilde{u}_i \tilde{u}_j\right)=\frac{\partial \bar{p}}{\partial x_i}+\frac{\partial}{\partial x_j}\left(\bar{\tau}_{i j}+\bar{\tau}_{i j}^t\right) \\
		\frac{\partial}{\partial t}\left(\bar{\rho} \tilde{Y}_s\right)+\frac{\partial}{\partial x_j}\left(\bar{\rho} \tilde{u}_j \tilde{Y}_s\right)=\frac{\partial}{\partial x_j}\left(\bar{\rho} D \frac{\partial \tilde{Y}_s}{\partial x_j}\right)-\frac{\partial \bar{\tau}_s}{\partial x_j}+\bar{\omega}_s \quad s=1, \ldots, n s-1\\
		\frac{\partial}{\partial t}(\bar{\rho} \tilde{E})+\frac{\partial}{\partial x_j}\left(\bar{\rho} \tilde{u}_j \widetilde{H}\right)=\frac{\partial}{\partial x_j}\left[k \frac{\partial \widetilde{T}}{\partial x_j}+\sum_{s=1}^{n s}\left(\bar{\rho} D \tilde{h}_s \frac{\partial \tilde{Y}_s}{\partial x_j}\right)+\tilde{u}_i \bar{\tau}_{i j}+\bar{\tau}_{\bar{e}}^t\right]
	\end{gathered}
\end{equation}
where $\bar{\rho}$ is mixture density, $t$ is time, $\tilde{u}_j$ is velocity component in the direction of $j$ in Cartesian coordinate system,
$x_j$ is spatial coordinate in the direction of $j$, $\bar{p}$ is filtering pressure,
$\bar{\tau}_{ij}$ and $\bar{\tau}_{ij}^t $ for molecular and turbulent viscosity tensor respectively,
$\tilde {Y_s} $ for the first $s $ mass fraction of a component, $D$ for the diffusion coefficient,
$\bar{\tau}_s=\bar{\rho}\left(\widetilde{u_jY_s}-\tilde{u}_j\tilde{Y_s}\right)$ for modeled
scalar turbulence flow, $\bar{\omega}_s$ is the filtered chemical reaction source term of component $s$,
$E$ is the total energy, $H$ is the total enthalpy, $T$ is the temperature, $k$ is the thermal conductivity,
$n s$ is the number of components, $\tilde{h}_s$ is the enthalpy of component.

\subsection*{Derivation of turbulent combustion source term}

The probability density function of the scalar in the grid $\boldsymbol{\phi}=\left\{Y_1, Y_2, \ldots Y_{n s-1}, H\right\}$ is
\begin{equation}
	\wp(\pmb{\phi};\mathbf{x},t)=\alpha\delta\left(\pmb{\phi}^m-\pmb{\phi}\right)+\beta\delta\left(\pmb{\phi}^*-\pmb{\phi}\right)
\end{equation}
where
\begin{equation}
	\delta(x)=\left\{\begin{array}{ll}
		+\infty, & x=0      \\
		0,       & x \neq 0
	\end{array}, \quad \int_{-\infty}^{+\infty} \delta(x) d x=1\right.
\end{equation}
for a probibility distribution, $\int_{-\infty}^{+\infty}\wp(\pmb{\phi};\pmb{x},t)d\pmb{\phi}=1$, which means $\alpha + \beta = 1$,
thus the averaged filtered property $\widetilde{\boldsymbol{\phi}}=$ $\left\{\tilde{Y}_1, \ldots, \tilde{Y}_{n s-1}, \tilde{H}\right\}$ can be represented as
\begin{equation}
	\widetilde{\boldsymbol{\phi}}=\frac{1}{\bar{\rho}} \int \wp(\boldsymbol{\phi} ; \boldsymbol{x}, t) \rho \boldsymbol{\phi} d \boldsymbol{\phi}=\frac{1}{\bar{\rho}}\left(\alpha \rho^m \boldsymbol{\phi}^m+\beta \rho^* \boldsymbol{\phi}^*\right)
\end{equation}

Assuming the density variation within a grid is minimal, $\rho^{m}=\rho^{*}=\bar{\rho}$,
\begin{equation}
	\widetilde{\phi}(\mathbf{x},t)=\alpha\boldsymbol{\phi}^m(\mathbf{x},t)+\beta\boldsymbol{\phi}^*(\mathbf{x},t)
\end{equation}

Considering the turbulent mix time as $\tau_{mix}$, we can derive the corresponding control equation of mass fraction:
\begin{equation}
	\bar{\rho}\dfrac{DY^*}{Dt}=-\dfrac{\bar{\rho}\left(\boldsymbol{Y}^*-\bar{Y}\right)}{\tau_{\mathrm{mix}}}+\dot{\omega}_s\left(\bar{\rho},\boldsymbol{\phi}^*\right)
	\label{eq:IEM equation}
\end{equation}

The source term of the chemical reaction, ${\dot{\omega}_{s}(\rho,\phi)}$, can be obtained by integrating the reaction rates
\begin{equation}
	\begin{array}{c}{{\overline{{\dot{\omega}_{s}(\rho,\phi)}}=\int\int_{\rho}\int_{\phi}\wp(\phi)\dot{\omega}_{s}(\rho,\phi)d\phi d\rho}}       \\
		{{\approx\alpha\dot{\omega}_{s}\left(\bar{\rho},\phi^{m}\right)+\beta\dot{\omega}_{s}\left(\bar{\rho},\phi^{*}\right)}} \\\end{array}
	\label{eq:omega_s_average}
\end{equation}

In stable pulsating state regions, the mass and energy are in a local equilibrium state, 
meaning that the mass/energy fraction mixing and reaction rate undergo a quasi-static process. 
Consequently, we let the left-hand side of Eq. \ref{eq:IEM equation} be equal to 0.
\begin{equation}
	\dfrac{\bar{\rho}\left(\boldsymbol{Y}^*-\widetilde{Y}\right)}{\tau_{\mathrm{mix}}}\approx\dot{\omega}_{s}\left(\bar{\rho},\boldsymbol{\phi}^*\right)
	\label{eq:balanced IEM equation}
\end{equation}

Considering the chemical reaction time as $\tau_c$, we simplify chemical process using 1-order approximation, i.e.
\begin{equation}
	\dot{\omega}\left(\bar{\rho},\phi^l\right)\approx\dot{\omega}(\bar{\rho},\widetilde{\phi})- \frac{1}{\tau_c}\cdot\bar{\rho}\left(\boldsymbol{\phi}^*-\widetilde{\phi}\right)
	\label{eq:chemical_taylor}
\end{equation}

Subsititute Eq.\ref{eq:balanced IEM equation} into Eq.\ref{eq:chemical_taylor}
\begin{equation}
	\dot{\omega}\left(\bar{\rho},\phi^*\right)\approx\dot{\omega}(\bar{\rho},\widetilde{\phi})-\dfrac{1}{\tau_c}\cdot\tau_{mix}\dot{\omega}\left(\bar{\rho},\phi^*\right)
\end{equation}

\begin{equation}
	\dot{\omega}\left(\bar{\rho},\phi^{*}\right)\approx\frac{\tau_{c}}{\tau_{c}+\tau_{\text{mix}}}\dot{\omega}(\bar{\rho},\tilde{\phi})
	\label{eq:omega*}
\end{equation}

Subsititute Eq.\ref{eq:omega*} into Eq.\ref{eq:omega_s_average}, and let $\frac{\tau_{c}}{\tau_{c}+\tau_{\text{mix}}}=\kappa$
\begin{equation}
	\overline{\dot{\omega}_s(\rho,\phi)}\approx\alpha\dot{\omega}_s(\bar{\rho},\phi^m)+(1-\alpha)\kappa\dot{\omega}_s(\bar{\rho},\widetilde{\phi})
	\label{eq:omega2}
\end{equation}
where $\alpha$ represents the weight of the mixing state of scalars, which denotes the homogeneity of the reaction scalars in the subgrid. 
In this study, an attempt is made to dynamically model $\alpha$ values without introducing additional variables. 
The parameter $\alpha$ should exhibit the following properties: it should lie between 0 and 1, 
and have a positive correlation with premixed condition in the flow field. Thus, flame index (FI) is introduced to identify the premix condition of the flow field

\begin{equation}
	\operatorname{FI}=\frac{\tilde{\nabla Y}_{Fuel}\cdot\tilde{\nabla Y}_{ox}}{\left|\nabla\tilde{Y}_{Fuel}\right|\times\left|\nabla\tilde{Y}_{ox}\right|}
\end{equation}
where $\nabla$ denotes the Nabla operator, with $\nabla=\frac{\partial}{\partial x} i+\frac{\partial}{\partial y} j+\frac{\partial}{\partial z} k$, $\tilde{Y}_{Fuel}$ 
and $\tilde{Y}_{ox}$ represent the mass fractions of fuel and oxidant, respectively. 
When $\operatorname{FI} = 1$, the local flow field exhibits a fully premixed state; when $\operatorname{FI} = -1$, the local flow field is in a completely unpremixed state. 
After identifying the premixed and unpremixed states of the local flow field, 
we compute $\alpha$ using the evenness of the weighted mixing fraction (corresponding to unpremixed conditions) and the reaction process variables (corresponding to premixed conditions). 
Drawing inspiration from Kuron et al.'s \cite{kuron_mixing_2017} mixing time scales model, the final expression for $\alpha$ is modeled as follows:
\begin{equation}
	\alpha=\frac{1-\mathrm{FI}}{2}\left[1-\frac{\widetilde{Z^{\prime\prime2}}}{\tilde{Z}(1-\tilde{Z})}\right]+\frac{1+\mathrm{FI}}{2}\left[1-\frac{\widetilde{c^{\prime\prime2}}}{\tilde{c}(1-\tilde{c})}\right]
	\label{eq:alpha}
\end{equation}
where $Z$ is the mixing fraction, which represents the mixing of fuel and oxidizer in the local flow field, and $c$ is the variation of reaction process, 
which represents the degree of reaction. Because $\boldsymbol{\phi^{\mathrm{m}}}\approx\widetilde{\boldsymbol{\phi}},\dot{\omega}\left(\bar{\rho},\boldsymbol{\phi^{\mathrm{m}}}\right)\approx\dot{\omega}(\bar{\rho},\widetilde{\phi})$,
substitute into Eq. \ref{eq:omega2}, the closed average chemical reaction source term can be written as

\begin{equation}
	\bar{\omega}_s(\rho,\phi)\approx\alpha\dot{\omega}_s(\bar{\rho},\widetilde{\phi})+(1-\alpha)\kappa\dot{\omega}_s(\bar{\rho},\widetilde{\phi})
	\label{Eq:2-delta model1}
\end{equation}

The model comprises two alpha-weighted components: as $\alpha$ approaches zero, 
it converges to the PaSR model. When $\alpha$ approaches 1, 
the subgrid fluctuations of the components within the grid approach zero, 
which can be considered as a homogeneous reactor. In this case, 
the filtered reaction rate corresponds to the reaction rate obtained through analytical quantities. 
Another way to interpret this model is to rewrite Eq. \ref{Eq:2-delta model1} as follows:
\begin{equation}
	\bar{\omega}_s(\rho,\phi)\approx\alpha(1-\kappa)\dot{\omega}_s(\bar{\rho},\widetilde{\phi})+ \kappa\dot{\omega}_s(\bar{\rho},\widetilde{\phi})
	\label{Eq:2-delta model2}
\end{equation}
where the first term on the right-hand side of Eq. \ref{Eq:2-delta model2} represents the homogeneous reactor term taking into account premixing,
while the second term represents the original PaSR model term. 
Therefore, the essence of this model is to consider dynamically the chemical reaction source term under the premixing state on the basis of the PaSR model.
The variances of the mixed fraction and the reaction process variables $\widetilde{\phi^{\prime\prime2}}$ and  $\widetilde{c^{\prime\prime2}}$ in Eq.\ref{eq:alpha} need to be further modeled for numerical simulation.
The model proposed by Pierce et al. \cite{pierce_dynamic_1998} based on the sub-grid balance hypothesis is used in this study, 
which assumes that the local balance exists between the generation rate and dissipation of a sub-grid variance of a conserved quantity. 
For variable $\phi$, its Favre variance $\widetilde{\phi^{\prime\prime2}}$ is given by, 
\begin{equation}
\widetilde{\phi''^2}=C\Delta^2(\nabla\tilde{\phi})^2
\end{equation}
where $\Delta$ is the grid filter length of the LES, representing the sub-grid turbulence length scale, and $C$ is the model constant. To avoid non-physical variances, 
$\widetilde{\phi^{\prime\prime2}}$ is restricted by $0 \leq \widetilde{\phi^{\prime \prime 2}} \leq \tilde{\phi}(1-\tilde{\phi})$.
The coefficients $\tau_c$ and $\tau_{\text{mix}}$ in the model parameter $\kappa=\frac{\tau_c}{\tau_c+\tau_{\text {mix }}}$ are obtained using analytical methods. 
$\tau_c$ adopts the defining method proposed by Golovitchev et al.\cite{golovitchev2001detailed}
\begin{equation}
\tau_c=\sum\limits_{r=1}^{nr}\frac{c_{tot}}{\sum_{n=1}^{NS,RHS}\nu_{n,r}k_{f,r}}
\end{equation}
where $c_{tot}$ is the total concentration of the mixture, $r$ and $nr$ are the index and total number of reactions, 
$\nu_{n,r}$ represents the stoichiometric coefficient of the n-th component in the i-th reaction, 
and $k_{f,r}$ is the forward reaction rate coefficient. The value of $\tau_{\text{mix}}$ is calculated using 
\begin{equation}
	\tau_{mix}=C_{mix}\sqrt{\dfrac{\mu_{eff}}{\rho\epsilon}}
\end{equation}
where $\mu_{\text{eff}}$ represents the sum of the turbulent viscosity coefficient and the molecular viscosity coefficient, 
while $\varepsilon$ denotes the dissipation rate of turbulent kinetic energy, 
and $C_{\text{mix}}$ serves as a model constant.

\section*{NUMERICAL SETUP}
\begin{figure}[!ht]
	\centering
	\includegraphics[width=0.8\linewidth]{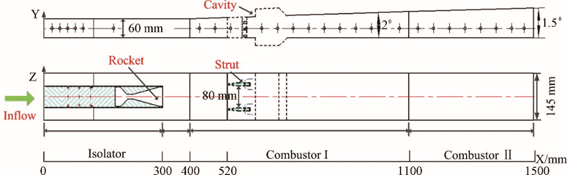}
	\caption{Diagram of RBCC model combutor.}
	\label{fig:diagram}
\end{figure}

To verify and evaluate the effectiveness of the new model, we implemented numerical simulation on a typical RBCC model combustor with a strut-cavity flameholder. Since it is difficult to organize the formation, 
ignition and combustion of the mixture under the condition of high speed incoming flow, the cross jet is usually adopted to enhance the oil-gas mixing, and the cavity or strut configuration is 
mostly adopted in the combustor to prolong the residence time of the gas, providing favorable conditions for ignition and stable combustion. 
The combustion chamber uses jet rocket for ignition, 
and uses both the strut and cavity to stabilize the flame, each side of the strut has 10 kerosene spray holes with a diameter of 0.5mm, and the fuel is ejected from the strut. The schematic diagram of RBCC is shown in Figure \Ref{fig:diagram}, and the 
inlet parameters for the isolator is shown in Table \ref{tab:inletParameters}.

\begin{table}[!ht]
	\begin{tabular}{ll}
	\hline
	Items                       & value \\ \hline
	Inlet Total Pressure (Mpa)  & 1.47  \\
	Inlet Total Temperature (K) & 1320  \\
	Inlet Mass Flow Rate (kg/s) & 3.93  \\
	Inlet Mach Number          & 2.3   \\ \hline
	\end{tabular}
	\centering
	\caption{Inlet parameters for the isolator}
	\label{tab:inletParameters}
	\end{table}

\begin{figure}[!ht]
	\centering
	\includegraphics[width=0.5\linewidth]{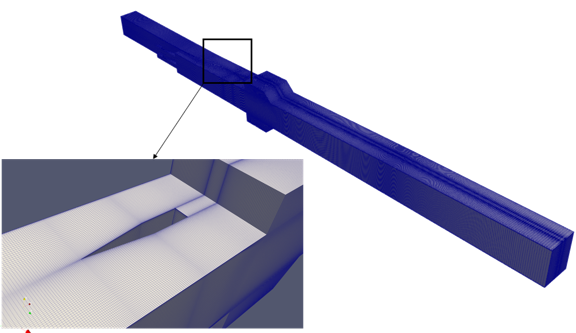}
	\caption{Mesh of the computational domain.}
	\label{fig:mesh}
\end{figure}

In this study, the mass flow rate of the rocket is 0.12kg/s, the equivalent ratio is 1.06, 
and the combustion chamber pressure is 1.57Mpa. According to the calculation of chemical equilibrium thermodynamics, 
the exit temperature of the rocket is 1855K. The exit group is divided into H2 (mass fraction of 0.06) and CO (mass fraction of 0.94).
And the kerosene flow from the strut is 0.2kg/s. For more detailed dimensions and boundary condition parameters of the configuration, 
please refer to \cite{xue_experimental_2017}.
In this study, large eddy simulation is used to simulate the configuration. The grids are full hexahedral, with a total of 12.0 million, as shown in Figure \ref{fig:mesh}.
The grids near fuel injection position, the strut region and the cavity are densed to capture the large-scale shear vortices in the flow field. 
A symmetric boundary condition is applied to the symmetry surface to save computational resource.
The solver uses the sprayFOAM solver in OpenFOAM 2.3.1, which can be used to solve chemical reaction simulations with droplet sprays. The turbulence model is one-equation k model \cite{kim_new_1995}. 
The combustion model adopts the original PaSR model and the dynamic model in this study. The adopted droplet breaking model is ReitzDiwakar model, and the droplet heat transfer model is RanzMarshall model.
After the flow field is stabilized, 
an average of five throughflows is computed as the mean flow field. 
The calculation takes approximately four days on a 10-node cluster with 24 cores each, 
resulting in a total computation time of around $2.3 \times 10^4$ CPU hours.

\section*{RESULTS AND DISCUSSION}
\begin{figure}[!ht]
	\centering
	\includegraphics[width=0.5\linewidth]{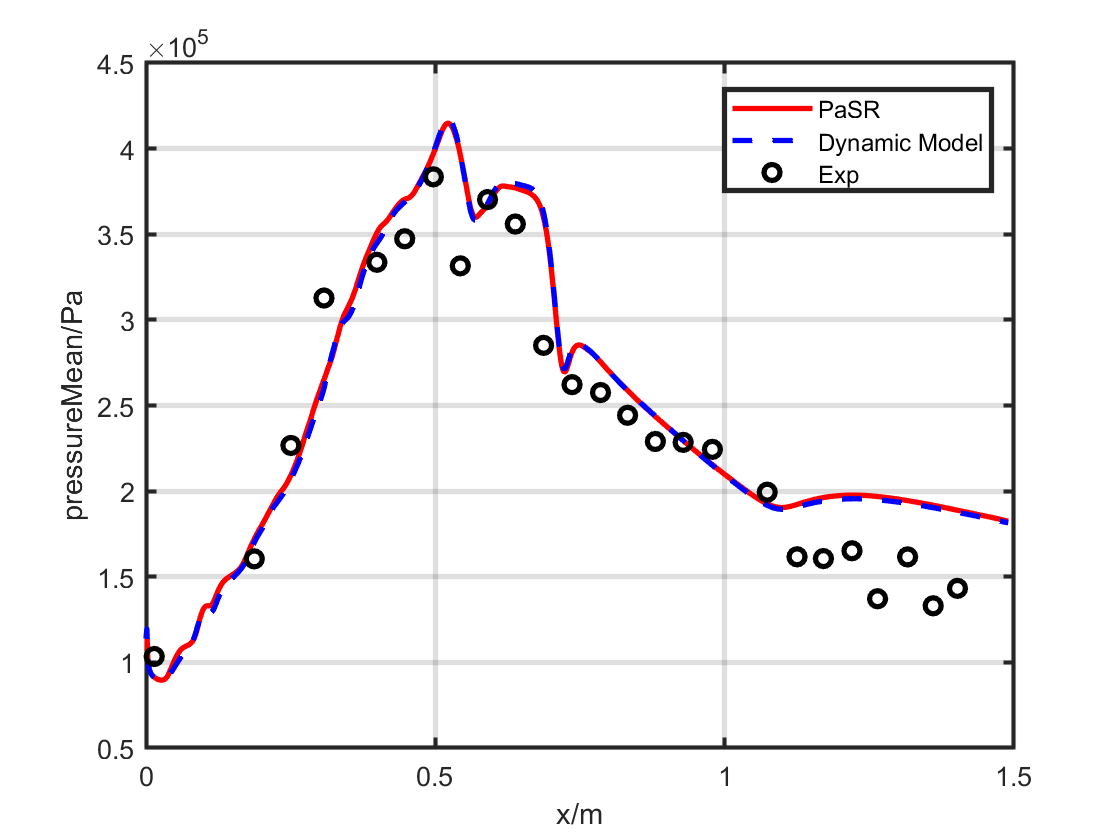}
	\caption{Comparison of static wall pressure between numerical results and experiments along streamwise direction.}
	\label{fig:expPressure}
\end{figure}

Figure \ref{fig:expPressure} presents a comparison of simulation and test results regarding static wall pressure along streamwise direction. 
As illustrated in the figure, 
both results corresponding to the two combustion models effectively capture the pressure changes of the hypersonic fluid within the combustion chamber. 
The static pressure increases from the inlet section, mainly due to the high back pressure generated by combustion, 
which results in the formation of a series of shock waves in the isolator section. 
As the airflow approaches the strut, the pressure near the strut first decreases and then increases, 
attributable to the choking effect induced by the strut itself. Owing to the thermal blockage created by high-temperature gas at the cavity's end, 
the airflow velocity increases, causing the combustor's pressure to gradually decrease after passing 
through the cavity. The static pressure exhibits minimal differences between the two models, necessitating 
further analysis of the internal flow field for comparison.

\begin{figure}[!ht]
	\centering
	\includegraphics[width=1\linewidth]{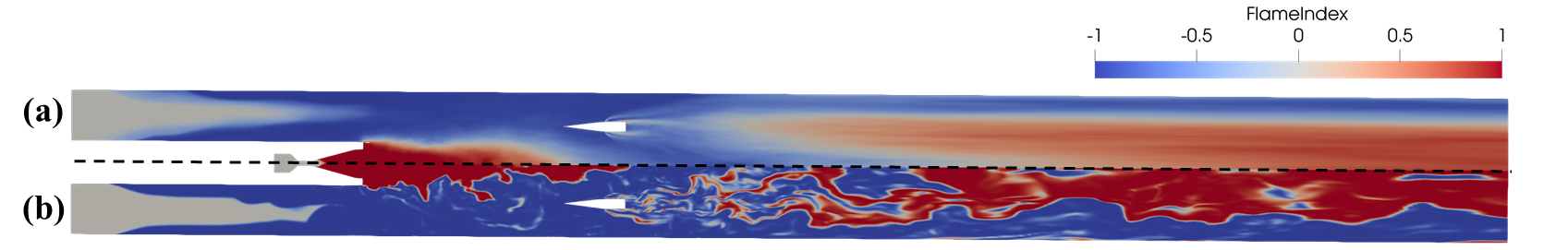}
	\caption{Average (a) and instantaneous (b) flame index field in strut view.}
	\label{fig:flameIndexStrut}
\end{figure}

To investigate the turbulent combustion mode within the combustion chamber, we further analyze the flow field information in the strut and cavity regions. 
Figure \ref{fig:flameIndexStrut} displays the average (a) and the instantaneous (b) flame index flow field of the strut view, 
respectively. From the instantaneous flow field, the flame index distribution exhibits a complex fragmented state. 
The jet flow at the rocket nozzle outlet mainly involves premixed combustion, while strong non-premixed combustion is primarily 
observed near the strut. As the reaction mixing layer thickens and the mixing between the strut and cavity intensifies downstream, 
the combustion gradually demonstrates strong premixed characteristics. 
Consequently, adopting a non-premixed/partially premixed dynamic combustion model becomes essential for a combustion chamber with a strut.


\begin{figure}[!ht]
	\centering
	\includegraphics[width=1\linewidth]{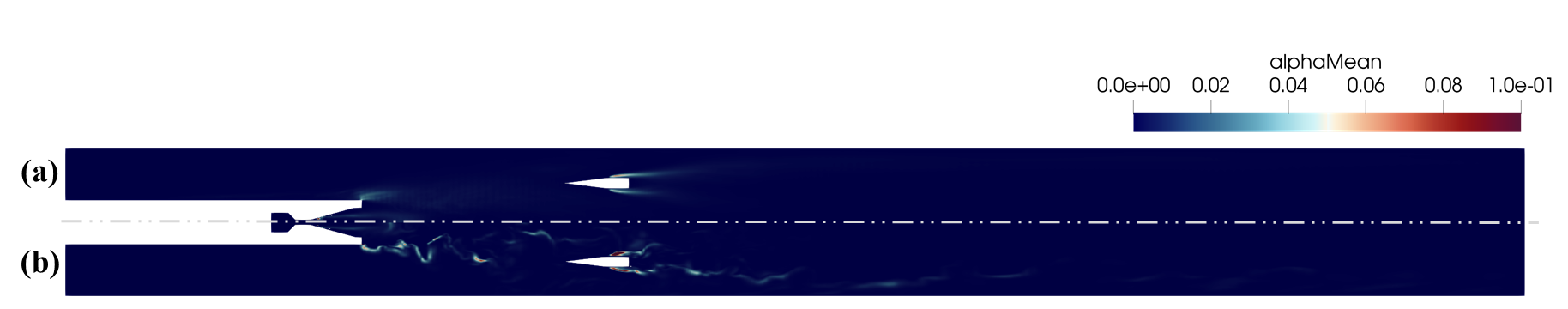}
	\caption{Average (a) and instantaneous (b) $\alpha$ field in strut view}
	\label{fig:alphaStrut}
\end{figure}

To analyze the effect of the proposed method in the premixed region, Figure \ref{fig:alphaStrut} displays the $\alpha$ distribution in the strut view, 
where (a) represents the average $\alpha$ flow field and (b) represents the instantaneous flow field at the end of the fifth through flow. 
From the figure, it is evident that the $\alpha$ values are primarily distributed at the tail of the jet flow at the rocket nozzle outlet 
and on both sides of the strut, predominantly governed by the flow of the wake. Although the flow field leans towards a strong premixed state 
at the tail of the combustion chamber, the $\alpha$ value remains small, as it is also associated with the process variables of the chemical 
reaction and their variance. In the central side of the tail of the combustion chamber, where combustion is approaching completion, 
the value of $\alpha$ is still not pronounced.

The specific differences in heat release rate between the new model and the dynamic model are compared to analyze the contribution of premixed combustion to the overall heat release.

\begin{figure}[!ht]
	\centering
	\includegraphics[width=1\linewidth]{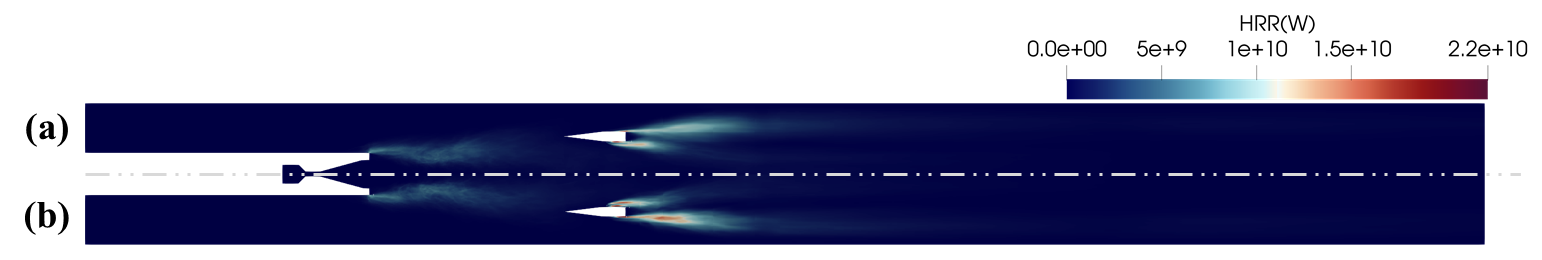}
	\caption{Heat release rate in strut view: PaSR model (a) and dynamic model(b)}
	\label{fig:dQStrut}
\end{figure}

Noting that Eq.\ref{Eq:2-delta model2} shows that the dynamic model additionally considers the chemical reaction source term under the premixing state, 
Figure \ref{fig:dQStrut} shows the heat release rate in the strut view, where (a) shows the heat release rate of the original PaSR model, 
and (b) shows the heat release rate result of the dynamic model. It can be seen that for this combustion chamber,
there is chemical reaction throughout the entire flow channel, and the dominant combustion region is located behind the strut. 
The estimate of the heat release rate of the new model near the strut is higher than that of the original PaSR model, 
which is consistent with the analysis result of the $\alpha$ distribution in Figure \ref{fig:alphaStrut}. 
The maximum heat release rate in the wake region increased from 1.2e10W to 1.6e10W. 
Therefore, the premixed chemical reacting source term predicted in this case is quite significant compared with the original PaSR.

\begin{figure}[!ht]
	\centering
	\includegraphics[width=0.9\linewidth]{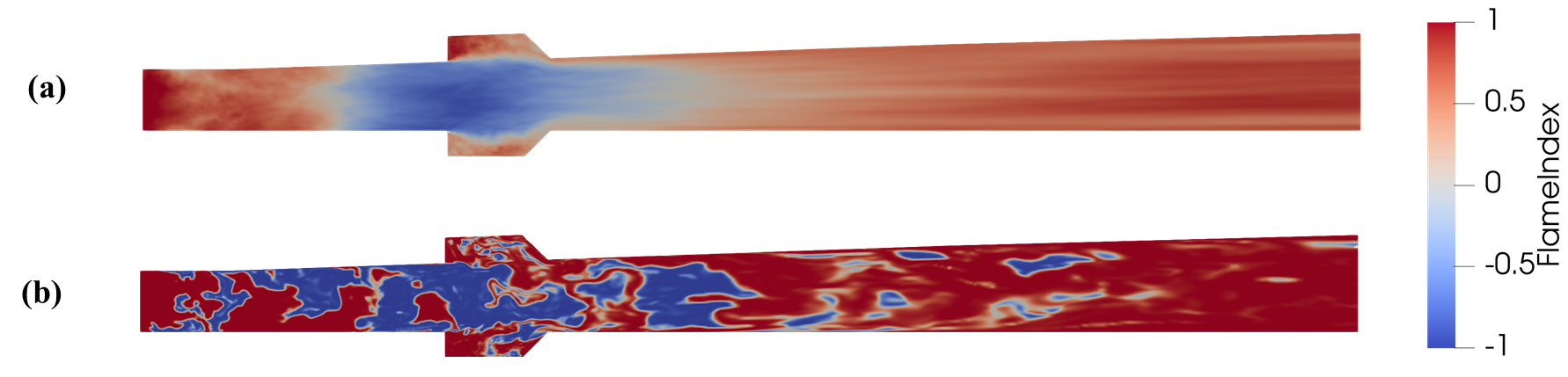}
	\caption{Average (a) and instantaneous (b) flame index in cavity view.}
	\label{fig:flameIndexCavity}
\end{figure}

Similar effort is made to examine the turbulent combustion mode in the cavity region and assess the improvements brought about 
by the new model on relevant parameters, reaffirming the need for dynamic modeling.
Figure \ref{fig:flameIndexCavity} displays the flame index distribution 
in the cavity view, with (a) representing the average flame index field and (b) illustrating the instantaneous flame index field results. The instantaneous flow field reveals that the area between the strut and cavity maintains complex premixed and 
non-premixed structural characteristics. However, based on the averaged flow field, the region where the strut is situated 
primarily demonstrates strong non-premixed features, while the region in the cavity predominantly exhibits strong premixed traits. 
As the chemical reaction layer thickens downstream, the downstream region increasingly showcases prominent premixed characteristics.

\begin{figure}[!ht]
	\centering
	\includegraphics[width=0.9\linewidth]{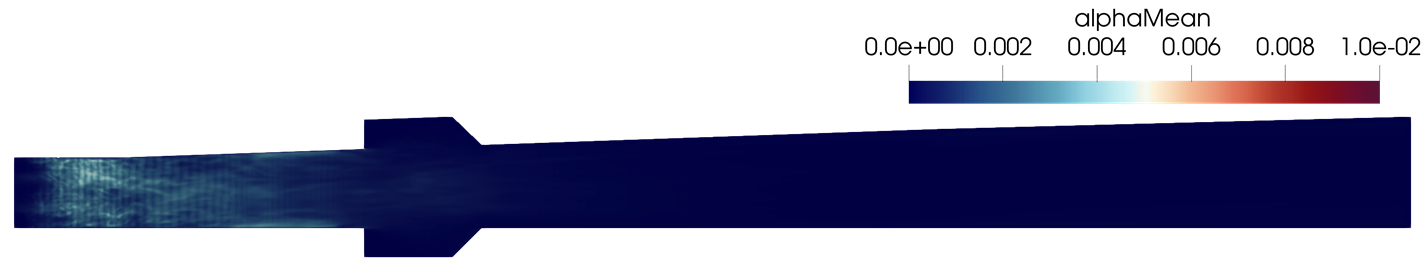}
	\caption{Mean $\alpha$ in cavity view.}
	\label{fig:alphaMeanCavity}
\end{figure}

While the premixed characteristics at the cavity are more pronounced than those at the strut, 
the $\alpha$ distribution in the cavity is not as prominent as that at the strut, 
as illustrated in Figure \ref{fig:alphaMeanCavity}. This is because, despite the larger flame index in this region, 
the reaction has already been much more complete at the strut. Consequently, 
the airflow entering the cavity exhibits a small variance in the chemical reaction process variable, 
resulting in a smaller $\alpha$ value, and the chemical reaction source term generated by premixing is not significant.


\section*{CONCLUSIONS}
In this study, we develop a partially premixed dynamic turbulence combustion model under the large eddy simulation 
framework and implement the model in the OpenFOAM code. The model is validated through simulations of a typical 
RBCC engine configuration, and the simulation results are compared with experimental data for static pressure, 
verifying the reliability of the simulation. Through analysis of the model, we obtain the following main conclusions:
\begin{enumerate}
	\item The RBCC engine model operates under a complex premixed and non-premixed combustion mode, 
which can be divided into three combustion zones. The first part is the reaction mixing layer formed by the high-temperature rocket exhaust jet and incoming air, 
which generates a large amount of heat and is characterized by premixed combustion. 
The second part is the violent mixing combustion of the fuel injection from the fuel strut and the reaction mixing layer formed by the rocket exhaust jet and incoming air, 
which releases a large amount of heat from the kerosene in a short distance, 
characterized by non-premixed combustion. The third part is the cavity region and a small amount of kerosene continuously burning and releasing a small amount of heat in the airflow downstream, 
playing an auxiliary role in stabilizing the flame, and characterized by premixed combustion.
	\item The new combustion model can dynamicly capture the premixed combustion situation. 
	The model returns to the PaSR model under completely non-premixed conditions, 
	and returns to the homogeneous reactor model under completely premixed homogenous conditions. 
	For the RBCC engine simulation example, the new combustion model mainly captures the premixed combustion situation in the strut and wake regions. 
	The prediction of the maximum heat release rate in the wake region is 30\% higher than that of the original PaSR model.
\end{enumerate}

\bibliographystyle{unsrt}

\bibliography{BibFiles/bibliography}

\end{document}